# Topological Phase Transition in a Magnetic Weyl Semimetal

D. F. Liu[1,2,*], Q. N. Xu[3,*], E. K. Liu[4,5,*], J. L. Shen[4], C. C. Le[3], Y. W. Li[6], D. Pei[6], A. J. Liang[2,7], P. Dudin[8], T. K. Kim[8], C. Cacho[8], Y. F. Xu[1], Y. Sun[3], L. X. Yang[9,10], Z. K. Liu[2,7], C. Felser[3,11,12], S. S. P. Parkin[1], Y. L. Chen[2,6,7,9,†]

[1]*Max Planck Institute of Microstructure Physics, Halle, 06120, Germany*
[2]*School of Physical Science and Technology, ShanghaiTech University, Shanghai, 201210, China*
[3]*Max Planck Institute for Chemical Physics of Solids, Dresden, D-01187, Germany*
[4]*Institute of Physics, Chinese Academy of Sciences, Beijing, 100190, China*
[5]*Songshan Lake Materials Laboratory, Dongguan, Guangdong, 523808, China*
[6]*Clarendon Laboratory, Department of Physics, University of Oxford, Oxford OX1 3PU, U.K.*
[7]*ShanghaiTech Laboratory for Topological Physics, Shanghai 200031, P. R. China*
[8]*Diamond Light Source, Didcot, OX110DE, U.K.*
[9]*State Key Laboratory of Low Dimensional Quantum Physics, Department of Physics, Tsinghua University, Beijing 100084, China*
[10]*Frontier Science Center for Quantum Information, Beijing 100084, China*
[11]*John A. Paulson School of Engineering and Applied Sciences, Harvard University, Cambridge, MA 02138, USA*
[12]*Department of Physics, Harvard University, Cambridge, MA 02138, USA*

\* These authors contributed equally to this work
† Email: yulin.chen@physics.ox.ac.uk

**Topological Weyl semimetals (TWSs) are exotic crystals possessing emergent relativistic Weyl fermions connected by unique surface Fermi-arcs (SFAs) in their electronic structures. To realize the TWS state, certain symmetry (such as the inversion or time reversal symmetry) must be broken, leading to a topological phase transition (TPT). Despite the great importance in understanding the formation of TWSs and their unusual properties, direct observation of such a TPT has been challenging. Here, using a recently discovered magnetic TWS $Co_3Sn_2S_2$, we were able to systematically study its TPT with detailed temperature dependence of the electronic structures by angle-resolved photoemission spectroscopy. The TPT with drastic band structures evolution was clearly observed across the Curie temperature ($T_C$ = 177 K), including the disappearance of the characteristic SFAs and the recombination of the spin-split bands that leads to the annihilation of Weyl points with opposite chirality. These results not only reveal important insights on the interplay between the magnetism and band topology in TWSs, but also provide a new method to control their exotic physical properties.**

# I. INTRODUCTION

Topological Weyl semimetals (TWSs) are novel topological quantum materials recently discovered with emergent relativistic bulk Weyl fermions connected by unique topological surface Fermi-arcs (SFAs) [1,2]. The unique electronic structures of TWSs not only give rise to rich unusual physical phenomena [1,3,4] and inspire theoretical discoveries [5,6], but also provide a platform for promising new applications such as high-speed electronics and optoelectronics due to the ultrahigh electron mobility and large magnetoresistance in TWSs [7,8], as well as the interplay between the topological SFAs and bulk Weyl fermions [9,10].

Unlike topological insulators (TIs) or topological Dirac semimetals, to realize a TWS, certain symmetry must be broken, such as the inversion symmetry (IS) [11-18] or the time reversal symmetry (TRS) [19-27], or both [28]. Such symmetry breaking leads to a topological phase transition (TPT) between the TWS and its neighboring states, as indicated by the change of topological invariants which can be reflected by a drastic change in the bulk and surface electronic structures, such as the emergence/disappearance of characteristic SFAs and the formation/annihilation of Weyl points (WPs) due to the recombination of the spin-split bulk bands. Therefore, investigation of such TPTs will not only help to understand the underlying mechanism for the formation of TWSs, but also give a new method to control the electronic structures and the exotic physical properties of these novel topological materials.

However, up to date, despite the extensive effort in the field, a direct observation on the evolution of electronic structures across a TPT between a TWS and its neighboring states has been challenging. As most of the TWSs discovered so far break the IS [11-18], an *in-situ* breaking/restoration of IS during the electron spectroscopic measurements would be desired, which is difficult to implement in experiments such as angle-resolved photoemission spectroscope (ARPES) which is carried out in ultrahigh vacuum (typically ~ $10^{-11}$ Torr) and requires the exposure of the sample surface.

Fortunately, the recent discovery of a magnetic TWS $Co_3Sn_2S_2$ [22-26] provides a new opportunity for the study of such a TPT. As illustrated in Fig. 1(a), below its Curie temperature ($T_C$ = 177 K), $Co_3Sn_2S_2$ enters the ferromagnetic (FM) TWS state which breaks the TRS; while above $T_C$, the restoration of the TRS causes the recombination of the two spin-split bands [which form two sets of WPs at different energies, see Fig. 1(a)], leading to the annihilation of WPs and the opening of a gap [29] (see Part III in Supplementary Information).

To investigate this TPT, we used ARPES to directly visualize the band structure of $Co_3Sn_2S_2$ and measure its temperature evolution across $T_C$. Indeed, we observed the characteristic SFAs in the TWS state that vanishes across $T_C$; furthermore, the separation of the two spin-split bands (that form the two sets of WPs) decreases monotonically with the increase of temperature, and the two bands eventually merged at $T_C$. These experimental discoveries, consistent with our ab initio calculations, reveal the drastic band structure change across the TPT in a TWS – which can provide important insights for the interplay between magnetism and topology, as well as the understanding of exotic physical properties in TRS-broken TWSs, such as the temperature dependence of the anomalous Hall conductivity (AHC) [22,23], anomalous Hall angle (AHA) [22] and anomalous Nernst effect (ANE) [30,31].

## II. RESULTS

In the TWS state of $Co_3Sn_2S_2$ [24,25], the WPs locate at the vicinity of the $\overline{M}$ and $\overline{M}'$ points along the boundary of the surface Brillouin zone (BZ) and are connected by the SFAs, forming characteristic triangle-shaped Fermi-surfaces (FSs) as illustrated in Fig. 1(b)(i). The FM order in the TWS state lifts the bulk bands' spin-degeneracy, forming two spin-split bands and two sets of WPs separated by $\Delta E = \sim 0.2$ eV according to our ab initio calculations [Fig. 1(b)(i), see Part IV in Supplementary Information]; on the other hand, the electronic structure of paramagnetic (PM) state for $T > T_C$ is illustrated in Fig. 1(b)(ii), showing no SFAs or WPs. Remarkably, the ab initio calculations (see Part III in Supplementary Information) show that the band inversion in $Co_3Sn_2S_2$ survives even when $T > T_C$, leading to the formation of a strong TI phase if the Fermi level lies in the bulk gap [29] (see Part III, VI and VII in Supplementary Information). For direct comparison, in Fig. 1(c) we plot the bulk band dispersions with the WPs in the FM state and the gapped bulk bands in the PM state; and Fig. 1(d) plots the FSs above and below $T_C$ that also show clear differences such as the disappearance of the SFAs and the obvious change of the bulk band's FS.

From ARPES measurements, the overall band structure of $Co_3Sn_2S_2$ below and above $T_C$ are summarized in Fig. 2. Clearly, the dramatic changes in the electronic structures are manifested by both differences on the topology of the constant energy contours [Figs. 2(a) and 2(b)] and the band dispersions across the whole BZ [Figs. 2(c) and 2(d)]. As an example, the FSs at $T = 6$ K and $T = 220$ K are plotted side by side in Fig. 2(b) to show that the SFAs near the $\overline{K}/\overline{K}'$ points [Fig. 2(b)(i)] disappear at $T = 220$ K [Fig. 2(b)(ii)] as well as the obvious differences between the bulk bands' FSs near the $\overline{\Gamma}$ and $\overline{M}/\overline{M}'$ points. Similarly, the experimental and theoretical band dispersions in Figs. 2(c) and 2(d) also illustrate clear differences below and above $T_C$.

To do a more detailed investigation, we first examine the evolution of SFAs with temperature, as illustrated in Fig. 3(a). At a temperature well below $T_C$ [Fig. 3(a)(ii), $T$ = 6 K], the SFAs form a large triangle-shaped FS around the $\overline{K}'$ point, consistent with the calculations in Fig. 3(a)(i) and the previous work [25]; when the temperature approaches $T_C$, the size of the triangle-shaped SFAs gradually decreases, and eventually disappears when $T > T_C$ [Fig. 3(a)(vii)], leaving only the bulk FS pocket, as supported by the calculations [Fig. 3(a)(viii)].

In addition to the temperature evolution of FSs, the evolution of the dispersions of topological surface states [TSSs, which form the SFAs at the Fermi energy ($E_F$)] are also obvious, as can be seen in Fig. 3(b): while deep in the FM state ($T$ = 6 K), the TSS can be clearly observed [Fig. 3(b)(ii)] in agreement with the calculations [Fig. 3(b)(i)]; with the increase of temperature, the TSS gradually shifts positions, changes the shape and eventually disappears when $T > T_C$ [Fig. 3(b)(vii)] (see Part VI and VIII in Supplementary Information), as confirmed by the calculations [Fig. 3(b)(viii)].

In order to check if the temperature evolution of the SFAs and the TSSs observed above are genuine, we cycled the sample temperature. As expected, when the temperature is cooled back down to below $T_C$, both the SFAs and the TSSs reemerge (see Part IX in Supplementary Information), confirming both evolutions as intrinsic.

Besides SFAs and TSSs, the bulk electronic structures also undergo dramatic change through the TPT: the separation of the spin-split Weyl bands in the FM (TWS) states will decrease with the elevation of temperature, and the two bands eventually recombine at $T = T_C$, leading to the annihilation of two sets of WPs with opposite chirality [Fig. 1(a)]. To investigate this process, we focus on the band dispersions that cut through the WPs [Fig. 4(a)] and their temperature evolution.

The ab initio calculations were first carried out to understand the band evolution through the TPT process. In the FM (TWS) state, the spin-up and spin-down bands are well separated, forming two sets of Weyl points with opposite chirality [Fig. 4(b)(i)]; while in the PM state, the two spin-split bands combine and annihilate the two WPs with the opening of a gap [see Fig. 4(b)(ii), and more discussion can be found in Part III and IV of Supplementary Information]. To track how the spin-split bands gradually merge, we computed their band positions [Fig. 4(c)(i)] and their separation [Fig. 4(c)(ii)] under different effective magnetic moments ($M_{eff}$) as a function of temperature (see Part X in Supplementary Information). As shown in Fig. 4(c) – the separation [Fig. 4(c)(ii)] reaches maximum of ~ 0.2 eV at low temperature ($T$ = 6 K, $M_{eff}$ = 0.3 $\mu_B$/Co), then decreases and eventually vanishes at $T_C$ (177 K, $M_{eff}$ = 0, see Part X in the Supplementary Information).

In ARPES measurements, we chose the 115 eV photon energy to access the Weyl band dispersions [25] (see Part IV and V in Supplementary Information) and monitor their evolution with temperature. As shown in Fig. 4(d), with the increase of temperature, the spin-up band (marked by red dashed lines) gradually moves up as a result of the decreased effective magnetic moment; while the spin-down band (marked by blue dashed lines) originally resides well above $E_F$ – thus cannot be seen by ARPES at low temperature – gradually moves down and can be observed when $T \geqslant$ 120 K [Figs. 4(d)(iii) and 4(d)(iv)].

When the temperature reaches $T_C$ (177 K), the two bands eventually merged together, which can be seen in Figs. 4(d)(v) and 4d(vi).

The experimental and calculated dispersions are plotted side by side in Fig. 4(e), showing nice agreements. For a more quantitative comparison, we extracted characteristic energy distribution curves (at $k_x = 0.4$ Å$^{-1}$) from the Weyl bands and assemble them together [Fig. 4(f)] to show the evolution of the two bands' position [Fig. 4(g)(i)] as well as their energy separation [Fig. 4(g)(ii)] as a function of temperature, which clearly show the gradual combination of the two bands, agreeing well with the calculated results in Fig. 4(c) and a recent optical study [32].

## III. CONCLUSION

The drastic temperature dependence of the band structures across $T_C$, including the disappearance of the characteristic SFAs and the recombination of the spin-split Weyl bands that leads to the annihilation of WPs with opposite chirality, together with the consistent theoretical calculations, revealed the mechanism of the temperature induced TPT in the magnetic TWS Co$_3$Sn$_2$S$_2$, which not only provides important insights on the interplay between the magnetism and band topology in TWSs, but also gives a new method to control their exotic physical properties, such as the temperature dependence of the AHC, AHA and ANE effects [22,23,30,31].

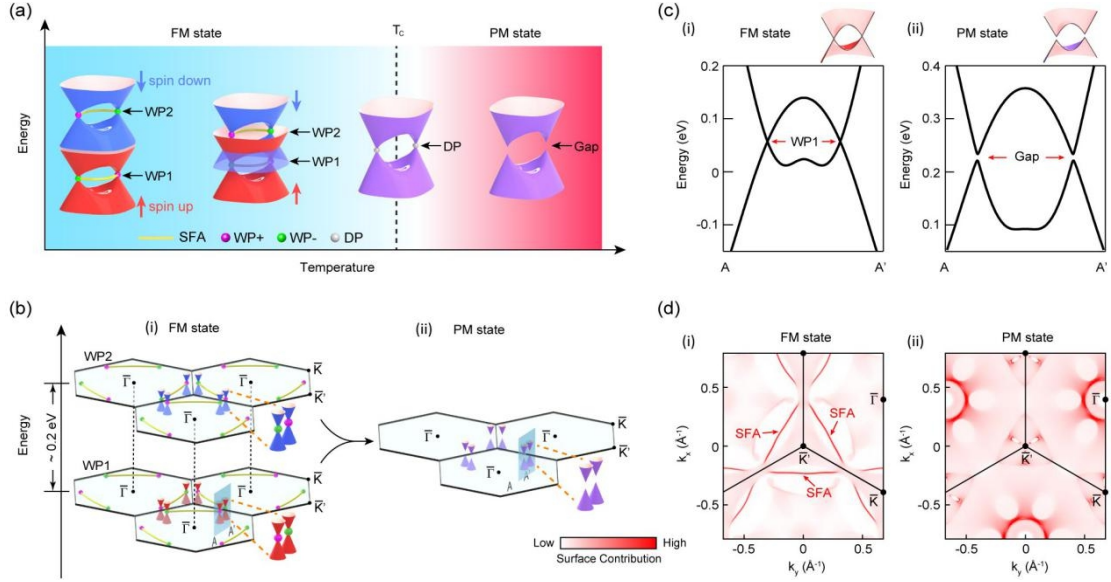

**Fig. 1. Topological phase transition across $T_C$ of $Co_3Sn_2S_2$.**

(a) Illustration of the topological phase transition with band structure evolution across $T_C$ (Curie temperature) in $Co_3Sn_2S_2$. Two sets of Weyl points at different energies are labeled as WP1 and WP2, respectively. Blue and red arrows indicate the spin-down and spin-up bands, respectively. Magenta and green color of the Weyl points represent positive (+) and negative (−) chirality, respectively. FM: ferromagnetism. PM: paramagnetism. DP: Dirac point. (b) Schematics of the bulk and surface electronic structure in the vicinity of Weyl points for the FM (i) and PM (ii) states. Bulk band dispersions across the WPs (i) and gapped bulk bands (ii) are also illustrated. Solid yellow lines connecting two WPs in (i) are surface Fermi-arc (SFAs). Note the SFAs disappear in the PM state in (ii). (c) Calculated bulk band dispersions across a pair of Weyl points [along AA' direction as indicated by the blue planes in (b)] in the FM state (i) and PM state (ii), respectively. An energy gap opens in the PM state, indicating a topological insulator phase. (d) Calculated Fermi surfaces from both bulk and surface states in the FM (i) and PM (ii) states. Note that the SFAs vanish in the PM state, as illustrated in (b).

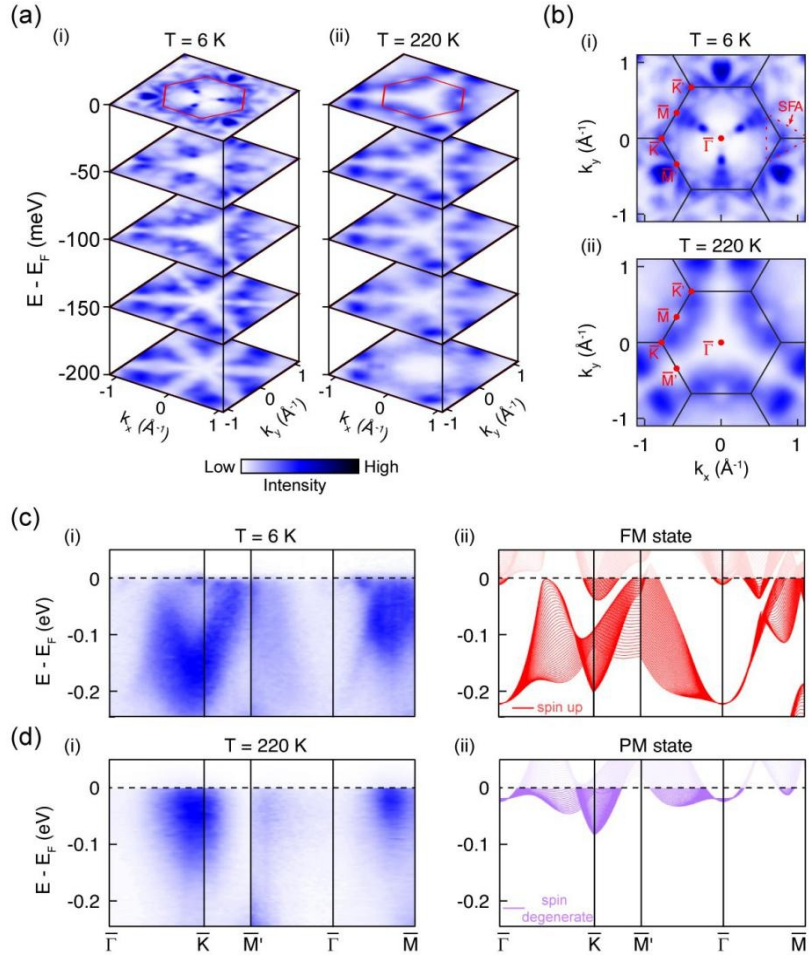

**Fig. 2. Comparison of the general electronic structure below and above $T_C$.**

(a) Stacking plot of constant energy contours at different binding energies below (i) and above (ii) $T_C$. The data has been symmetrized according to the crystal symmetry. (b) Comparison of the Fermi surface topology below (i) and above (ii) $T_C$. The triangle-shaped SFAs near the $\overline{K}/\overline{K}'$ points are clearly observed below $T_C$ (i), which vanish above $T_C$ (ii). (c) Comparison of the experimental (i) and calculated (ii) band dispersions in the FM state along different high-symmetry directions across the whole Brillouin zone (BZ), showing good agreement. Note the calculated bandwidth was renormalized by a factor of 1.43, and the energy position was shifted to match the experiment (same below). Data were collected using photons at 125 eV with linear horizontal (*LH*) polarization. (d) Same as (c) for the PM state, the experimental and calculated results again show good agreement.

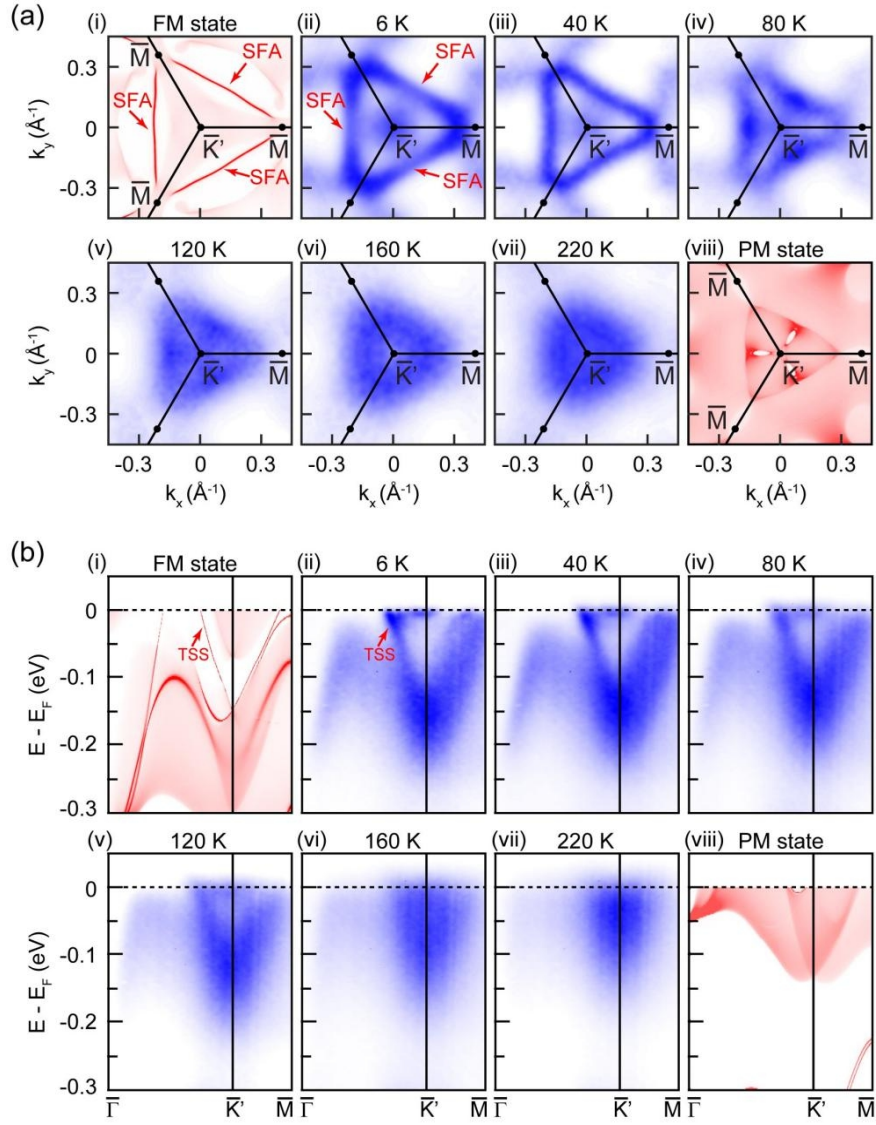

**Fig. 3. Temperature evolution of SFAs across $T_C$.**

(a) Comparison of the calculated Fermi surfaces of bulk and surface states in the FM (i) and PM (viii) states to the experimental results at different temperatures (ii)-(vii). The triangle-shaped SFAs [indicated by red arrows in (ii)] gradually shrink in size with increasing temperature and eventually vanish above $T_C$ (vii), leaving only the bulk Fermi surface pocket. The data has been symmetrized according to the crystal symmetry. Here we use all three $\bar{M}$ points in (i) and (viii) for simplicity as $\bar{M}/\bar{M}'$ points are equivalent for the surface states. (b) Comparison of the calculated dispersions of topological surface state (TSS) in the FM (i) and PM (viii) states to experimental results at different temperatures (ii)-(vii). The TSS (indicated by red arrows) gradually moves up and eventually disappear upon increasing temperature to above $T_C$, agreeing well with the calculations.

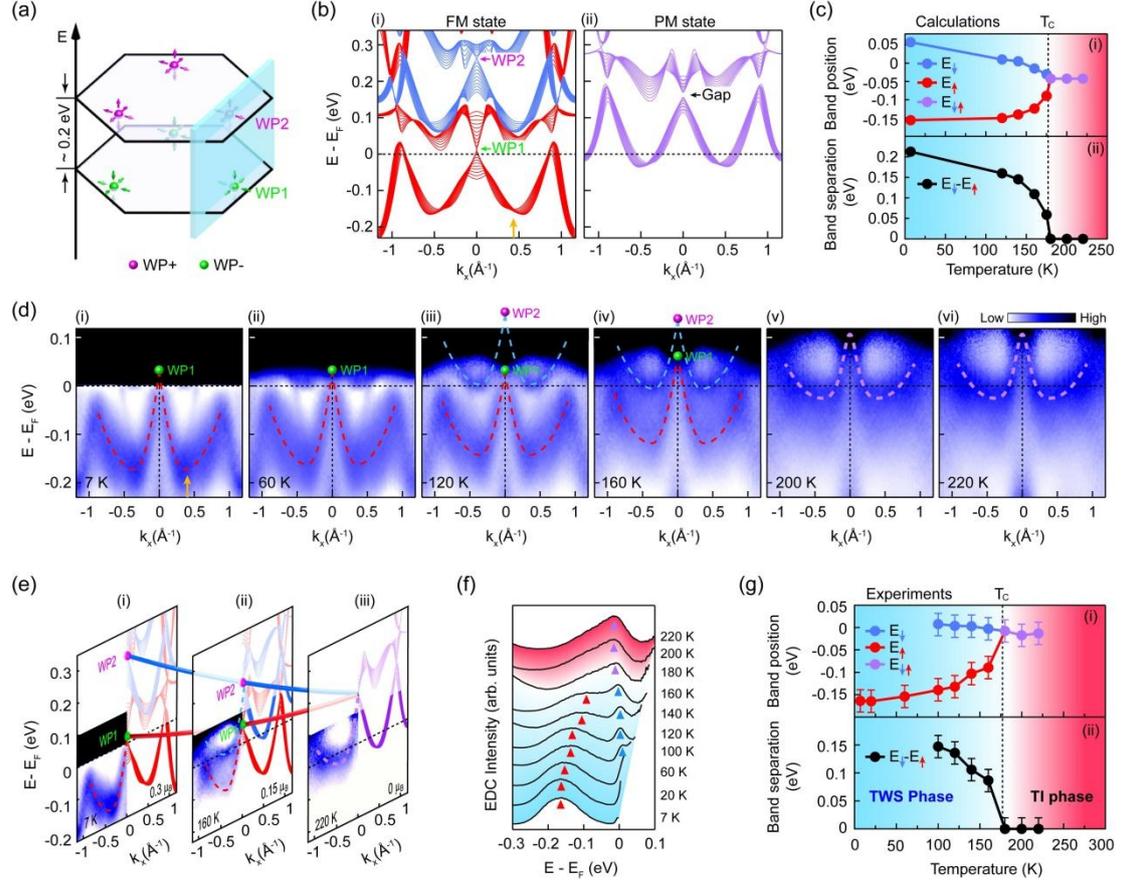

**Fig. 4. Temperature evolution of the Weyl band dispersions.**

(a) Schematic of the two sets of Weyl points from different spin-polarized bands with different energies. The blue plane indicates the dispersion direction of the bands illustrated in (b) and (d). Definition of WP1 and WP2 is the same as in Fig. 1. (b)(i) Calculated band dispersions along the plane indicated in (a), in the FM states. Red and blue curves represent the spin-up and spin-down bands, respectively; and the two Weyl points (WP1, WP2) are marked in both bands. (ii) The two spin-split bands in (i) merge in the PM sates (shown as magenta curves); with the two Weyl points merge and annihilate, forming an energy gap as indicated. (c) Calculated energy position [at the momentum indicated by the orange arrow in (b)(i)] for spin-up and spin-down bands (i) and their separation (ii) as a function of temperature. (d) Temperature dependence of the dispersion across the Weyl points. Red and blue dashed curves are the guidelines of two sets of Weyl dispersions. The definition of red, blue and magenta color is the same as in (b). Note that the experimental data were obtained through dividing the Fermi-Dirac function of the raw data to highlight the unoccupied state above $E_F$ and then symmetrization with respect to $k_x$=0 was applied. (e) Side-by-side comparison of the experimental results and the calculations. (f) Temperature evolution of the experimental energy distribution curves [EDCs, extracted from $k_x = 0.4$ Å$^{-1}$ as indicated by the orange arrow in (d)(i)]. The red, blue and magenta triangles mark the spin-up, spin-down and merged bands' energy loci. (g) Temperature dependent plot of the spin-up, spin-down and merged bands' energy loci (i) and their energy separation (ii), showing excellent agreement with the calculations in (c).